\documentclass[aps,prb,amsmath,reprint,superscriptaddress,longbibliography]{revtex4-1}
\usepackage{amsmath,txfonts,bm,color,dcolumn,graphicx}
\usepackage{listings}
\usepackage{xcolor}

\colorlet{punct}{red!60!black}
\definecolor{background}{HTML}{EEEEEE}
\definecolor{delim}{RGB}{20,105,176}
\colorlet{numb}{magenta!60!black}

\lstdefinelanguage{json}{
    basicstyle=\scriptsize\ttfamily,
    numbers=none,
    stepnumber=1,
    numbersep=8pt,
    showstringspaces=false,
    breaklines=true,
    frame=lines,
    backgroundcolor=\color{background},
    literate=
     *{0}{{{\color{numb}0}}}{1}
      {1}{{{\color{numb}1}}}{1}
      {2}{{{\color{numb}2}}}{1}
      {3}{{{\color{numb}3}}}{1}
      {4}{{{\color{numb}4}}}{1}
      {5}{{{\color{numb}5}}}{1}
      {6}{{{\color{numb}6}}}{1}
      {7}{{{\color{numb}7}}}{1}
      {8}{{{\color{numb}8}}}{1}
      {9}{{{\color{numb}9}}}{1}
      {:}{{{\color{punct}{:}}}}{1}
      {,}{{{\color{punct}{,}}}}{1}
      {\{}{{{\color{delim}{\{}}}}{1}
      {\}}{{{\color{delim}{\}}}}}{1}
      {[}{{{\color{delim}{[}}}}{1}
      {]}{{{\color{delim}{]}}}}{1},
}

\begin{document}

\title{BAGEL: Brilliantly Advanced General Electronic-structure Library}
\author{Toru Shiozaki}
\email{shiozaki@northwestern.edu}
\affiliation{Department of Chemistry, Northwestern University, 2145 Sheridan Rd., Evanston, IL 60208, USA.}
\date{\today}

\begin{abstract}
On behalf of the development team, I review the capabilities of the BAGEL program package in this article.
BAGEL is a newly-developed full-fledged program package for electronic-structure computation in quantum chemistry,
which is released under the GNU General Public License with many contributions from the developers.
The unique features include analytical CASPT2 nuclear energy gradients and derivative couplings,
relativistic multireference wave functions based on the Dirac equation,
and implementations of novel electronic structure theories. 
All of the programs are efficiently parallelized using both threads and MPI processes.
We also discuss the code generator SMITH3, which has been used to implement some of the programs in BAGEL. 
The developers' contributions are listed at the end of the main text. 
\\
{\color[rgb]{0.6,0.6,0.6}\it Invited Software Focus article, Wiley Interdisciplinary Reviews: Computational Molecular Science}
\end{abstract}

\maketitle

\section{Introduction}
On August 22nd, 2012,
my friend Edward Valeev and I were having lovely cups of coffee at Mill Mountain Coffee \& Tea in Downtown Roanoke.
Our conversation was about what to name the electronic structure program package\cite{bagel} that my then brand new research group at Northwestern
had begun developing; it was there that I accepted his suggestion that we should name the package BAGEL.
\footnote{The first occurrence of the word BAGEL was in the commit message
          for Libint on August 23rd, 2012, which read ``Added test code to compare with ToruS' Bagel integrals\protect{"}}
The only group member at the time, Matthew MacLeod, later proposed that we brand BAGEL as an abbreviation for
{``Brilliantly Advanced Electronic-structure Library."}
This was the beginning of our brilliant journey, though
the roots of the program can be traced back to the integral code written in 2009 for explicitly correlated theories
with periodic boundary conditions.\cite{Shiozaki2009CPL,Shiozaki2010JCP}
Since then, many collaborators have contributed to the development of the BAGEL package.

Our motivation for developing a new electronic-structure program package
has been partly the notion that the theory developers and end users may benefit from having software programs that are specifically designed for
modern parallel computer hardware.
For example, in many of the supercomputer facilities, large local hard disks on compute nodes are not available anymore
in favor of inter-node distributed memory.
In addition, the memory space per node has increased tremendously in the past decade, and nodes with 64 or 128 gigabytes of memory are now simply commodity (which means 
that the memory space per 64-node rack amounts to a few terabytes).
This increase in the size of available memory has dramatically changed the domain of what can be stored in memory.
For instance, we store three-index integrals in density fitting algorithms in memory, since the size of the three-index integrals
is smaller than typical memory spaces even for molecules consisting of 100 atoms or more. 
And last but not the least, this kind of hardware is coming to every research group in the near future:
the price of computer clusters with 64 nodes (128 CPUs) is cheaper than one femtosecond Ti-Sapphire laser in a spectroscopy laboratory.

Our BAGEL program package\cite{bagel} is entirely written in C++. While the choice of the programming language is not important,
we consider that the object-oriented design that the C++ language facilitates
does have affinity to modern parallel computing.
All of the programs in BAGEL are parallelized using both threads and MPI processes.
Threading is performed using intra-node task queues, whose backend is either C++11 std::thread or OpenMP.
Furthermore, some of the programs in the BAGEL package are implemented using the automatic code generator, SMITH3,\cite{smith}
which has allowed for implementing complex algebraic expressions into efficient computer code, such as those for CASPT2 nuclear energy gradient evaluation (see below).
The programs generated by SMITH3 are parallelized using the remote memory access protocol in the MPI3 standard.

BAGEL is a full-fledged stand-alone program package.
It is only dependent on the BLAS and Lapack/ScaLapack libraries and the BOOST library.
BOOST is mainly used for parsing input files written in the JSON or XML format and for serialization.
For BLAS and Lapack/ScaLapack, Intel MKL is strongly recommended as it provides near-optimal efficiency and several useful extensions such as matrix transposition.
BAGEL is strictly standard compliant and can be complied by any MPI C++ compiler that fully supports the C++11 and MPI3 standards.
BAGEL is open source and is distributed under the GNU General Public License Version 3+, which allows users to freely download, modify, and redistribute it under the same license.
The source code and documentation can be found on the web portal at http://www.nubakery.org.

\section{Program Features}
In the following, we summarize the features of the BAGEL program package. Note that {\it every algorithm is implemented using density fitting} (see also Sec.~\ref{dfsec}).

\subsection{Standard algorithms}
{\bf Single-reference models:} The basis of the efficient programs in BAGEL is its parallel Fock builder using density fitting.
The code is highly efficient: for 8 benzene molecules using the cc-pVTZ basis set and the corresponding JKFIT auxiliary basis set for density fitting
(2112 basis functions, 5232 auxiliary functions, and 336 electrons), the entire Hartree--Fock SCF procedure (11 iterations)
takes only about 2~min (i.e., 10~sec per iteration).
Note that the hardware used to measure this timing was purchased in 2012; the timing should be substantially better on more recent machines.
BAGEL implements RHF, UHF, and ROHF.
The pilot KS-DFT code is also implemented using Libxc,\cite{Marques2012CPC} but it is neither efficient nor of production quality;
we are hereby soliciting future collaboration for improving the DFT code. 
In addition, RHF based on the fast multipole methods (FMM) including exact exchange is implemented for systems larger than a few hundreds of atoms.\cite{Le2017JCTC}
MP2 is implemented as a post-HF method to account for dynamical electron correlation.
The programs for computing analytical nuclear gradients for RHF, UHF, ROHF, and MP2 are available, all of which are efficiently parallelized.

{\bf Multireference models:} BAGEL implements CASSCF using a two-step second-order algorithm
in which density fitted integrals are directly contracted to trial vectors in each microiteration for orbital updates.
This algorithm is extremely robust and almost as cheap as the first-order algorithms
(note that each microiteration is cheaper than one Hartree--Fock iteration using density fitting).
As post-CASSCF methods, CASPT2, sc-NEVPT2, and MRCI are implemented.
Using the CASPT2 program we can also emulate pc-NEVPT2.
The multistate variants of CASPT2 (MS-CASPT2 and XMS-CASPT2) are also available.
The analytical nuclear gradients and derivative couplings for SA-CASSCF and XMS-CASPT2 are implemented as described in Sec.~\ref{caspt2grad}.
Furthermore, CASSCF in BAGEL can be run with external FCI solvers, in which active Hamiltonians and reduced density matrices are communicated between packages (i.e., loose coupling).
As an example, we developed an interface to the NECI program\cite{NECI,Booth2014MP} to enable CASSCF with very large active spaces using the FCI-QMC algorithm. 
This interface is extensible to other FCI solvers such as DMRG.\cite{Chan2011AnnRevPhys}

{\bf Geometry optimization and dynamics:} BAGEL optimizes molecular geometries using non-redundant or redundant internal coordinates.\cite{Schlegel2011Wiley}
The optimizer can be used for finding the geometries of equilibrium structures, transition states, and minimum energy conical intersections. 
In addition, the interfaces to dynamics programs, such as Newton-X\cite{newtonx} are available to allow for on-the-fly ab initio dynamics simulations using
the nuclear forces and derivative couplings computed by BAGEL.
Nuclear Hessian matrix elements, harmonic vibrational frequencies, and infrared absorption spectra
can also be computed from analytical nuclear gradients using the finite difference formula.\cite{Vlaisavljevich2017JCTC}
Finite difference calculations can be performed in the embarrassingly parallel mode.

{\bf Molden interface:}
BAGEL can generate output files in the Molden format\cite{molden} that contain molecular orbitals, vibrational frequencies, or geometry optimization information.
In addition, users can start BAGEL calculations using the information stored in the Molden file.
This feature is particularly useful when users have to impose spatial symmetry (as is sometimes the case for small symmetric molecules),
since the programs in BAGEL do not utilize any point-group symmetry at this point.
If imposing spatial symmetry is necessary, however, users can elect to generate and store
the geometry, basis functions, and molecular orbitals in the Molden format using other program packages
and start BAGEL calculations using them.

\begin{figure}
\begin{lstlisting}[language=json]
   { "bagel" : [ 
   {
     "title" : "molecule",
     "basis" : "svp",
     "df_basis" : "svp-jkfit",
     "geometry" : [ 
     { "atom" : "C", "xyz" : [ -0.07,  2.54,  0.00 ] },
     { "atom" : "C", "xyz" : [  2.55,  2.54,  0.00 ] },
     { "atom" : "C", "xyz" : [  3.87,  4.82,  0.00 ] },
     { "atom" : "C", "xyz" : [  2.55,  7.10,  0.00 ] },
     { "atom" : "C", "xyz" : [ -0.07,  7.10,  0.00 ] },
     { "atom" : "C", "xyz" : [ -1.39,  4.82,  0.00 ] },
     { "atom" : "H", "xyz" : [ -1.11,  0.74,  0.00 ] },
     { "atom" : "H", "xyz" : [  3.59,  0.74,  0.00 ] },
     { "atom" : "H", "xyz" : [  5.95,  4.82,  0.00 ] },
     { "atom" : "H", "xyz" : [  3.59,  8.90,  0.00 ] },
     { "atom" : "H", "xyz" : [ -1.11,  8.90,  0.00 ] },
     { "atom" : "H", "xyz" : [ -3.47,  4.82,  0.00 ] } 
     ]
   },
   {
     "title" : "optimize",
      "method" : [ { 
        "title" : "caspt2",
        "smith" : { 
          "method" : "caspt2",
          "shift" : 0.2,
          "frozen" : true
        },  
        "nact" : 6,
        "nclosed" : 18, 
        "active" : [17, 20, 21, 22, 23, 30] 
      } ] 
   }
   ]}
\end{lstlisting}
\caption{Sample input in the JSON format for the CASPT2 geometry optimization (benzene with the full $\pi$ active space).\label{figinput}}
\end{figure}

\subsection{CASPT2 nuclear gradients and derivative couplings\label{caspt2grad}}
Since the early 1990s, CASPT2 has been used in many chemical applications that involve multi-configurational electronic structure,
including those in photochemistry and inorganic chemistry.\cite{Roosbook,Pulay2011IJQC}
The first-order wave functions in CASPT2 are expanded in the so-called fully internally contracted basis functions, 
whose size scales only polynomially with respect to the number of active orbitals.
However, the analytical nuclear energy gradients for CASPT2 had not been realized mainly owing to the complexity of the equations and implementations associated with the full internal contraction,
until MacLeod and Shiozaki reported its implementation in the BAGEL package using an automatic code generation approach\cite{MacLeod2015JCP}
(note, however, an important contribution by Celani and Werner for a partially contracted variant\cite{Celani2003JCP}).
BAGEL currently implements nuclear energy gradients for multistate CASPT2 methods, MS-CASPT2 and XMS-CASPT2,\cite{Vlaisavljevich2016JCTC}
and derivative coupling between (X)MS-CASPT2 states.\cite{Park2017JCTC}
The computational cost of evaluating the nuclear gradients is typically only 2--3 times as expensive as that for the energy calculation.
The details on the timing and parallel performance can be found in Ref.~\onlinecite{Park2017JCTC2}.
Taking advantage of the efficiency, the program has been used in on-the-fly nonadiabatic dynamics studies.\cite{Park2017JCTC2}
The first-order properties such as dipole moments and hyperfine coupling constants can also be computed using the relaxed (spin) density matrices.\cite{Shiozaki2016JCTC}
The sample input is shown in Fig~\ref{figinput}.

\subsection{Relativistic multireference wave functions}
BAGEL implements various relativistic multireference wave function methods within the four-component Dirac formalism.\cite{Reiherbook}
The Dirac--Coulomb, Dirac--Gaunt, and Dirac--Breit interaction can be employed.\cite{Shiozaki2013JCP}
The program uses the density fitting approximation in the Fock build and MO transformation.\cite{Kelley2013JCP}
The four-component relativistic CASSCF implementation was first implemented using a quasi-second-order algorithm,\cite{Bates2015JCP}
which has subsequently been replaced by a program based on the second-order algorithm (unpublished).
The relativistic CASSCF program can perform minimax optimization for molecules with 100 atoms and a few heavy elements.
In CASSCF, the so-called quaternion diagonalization is used to impose the time-reversal symmetry,\cite{Shiozaki2017MP}
which is important for numerical stability.
The relativistic CASPT2 and MRCI methods have been implemented on the basis of the CASSCF program,\cite{Shiozaki2015JCTC}
which have recently been extended to multistate variants.\cite{Reynolds2017JCTC}
The EPR Hamiltonian parameters, such as zero-field splitting tensors, can be extracted from these computations.\cite{Reynolds2017JCTC}
Pilot implementations of nuclear energy gradients for mean-field models\cite{Shiozaki2013JCTC}
and programs for molecules under an external magnetic field using gauge-including atomic orbitals\cite{Reynolds2015PCCP}
are also available. 

\subsection{BAGEL as a theory development platform}
The compactness of the code in BAGEL allows us to efficiently implement new theory and algorithms as production code.
For instance, the so-called active space decomposition methods\cite{Parker2013JCP,Parker2014JPCC,Parker2014JCTC,Parker2014JCP,Kim2015JCTC} have been implemented in the BAGEL package.  
BAGEL has also been used as a development platform by researchers outside the BAGEL development team (see, for instance, Ref.~\onlinecite{Yao2016JCTC}).

\section{Implementation details} 
\subsection{Density fitting utilities\label{dfsec}}
The flexibility and compactness of the code in BAGEL mainly come from the design of its density fitting utilities,
which help manipulate three-index tensors $I_{\gamma,rs}$ in the density fitting algorithms.
These three-index tensors are kept in distributed memory.
The following operations are implemented in the density fitting class, and the code is efficiently parallelized
($M$, $I$, and $G$ are arbitrary two-, three-, and four-index tensors, respectively).
(1) Index transformation:
\begin{align}
&I'_{\gamma,ts} = \sum_{r} I_{\gamma,rs}M_{rt},\quad
I'_{\gamma,rt} = \sum_{s} I_{\gamma,rs}M_{st},\nonumber\\
&I'_{\gamma,rs} = \sum_{\delta} M'_{\gamma\delta} I_{\delta,rs},
\end{align}
which can be used for both MO transformations and AO back transformations;
(2) Contraction of two density fitting objects: 
\begin{align}
&M_{st} = \sum_{\gamma r} I_{\gamma,rs}I'_{\gamma,rt},\quad
G_{rs,tu} = \sum_\gamma I_{\gamma,rs}I'_{\delta,tu},\nonumber\\
&M_{\gamma\delta} = \sum_{rs} I_{\gamma,rs}I'_{\delta,rs};
\end{align}
(3) Contraction of density matrices and density fitting objects: 
\begin{align}
&D_{\gamma} = \sum_{rs} I_{\gamma,rs} M_{rs},\quad
M_{rs} = \sum_{\gamma} D_\gamma I_{\gamma,rs},\nonumber\\
&I'_{\gamma,tu} = \sum_{rs} I_{\gamma,rs} G_{rs,tu}.
\end{align}
We also implemented several specializations for the last operation when two-particle density matrices with closed orbitals are used as $G$.
Some of the basic level-1 BLAS-like operations (e.g., scaling, addition, and dot products) are also implemented. 
It is surprising that numerous quantum chemical algorithms with density fitting (HF, CASSCF, $Z$-vector equations, etc.) can be expressed in terms of these operations!

\subsection{Molecular integrals}
BAGEL implements all of the necessary molecular integrals using the standard algorithms.
The electron-repulsion integrals and their geometric derivatives are evaluated using the Rys-quadrature algorithm;\cite{Dupuis1976JCP}
the efficiency of our integral code is comparable to that of Libint by Valeev (one can optionally use Libint within BAGEL as well).\cite{libint}
In addition to the standard electron-repulsion integrals, BAGEL implements somewhat unique two-electron integrals, which are 
the two-electron spin--spin coupling integrals,\cite{Shiozaki2013JCP} the full Breit integrals in relativistic theory,\cite{Shiozaki2013JCP}
and the Slater-type geminal integrals for explicit correlation.\cite{Shiozaki2009CPL} 
The expressions for the integral kernel are shown below:
\begin{align}
\frac{1}{r_{12}},\quad \frac{3w_{12}w'_{12}-\delta_{ww'}r_{12}^2}{r_{12}^5},\quad \frac{w_{12}w'_{12}}{r_{12}^3},\quad e^{-\gamma r_{12}},\quad \frac{e^{-\gamma r_{12}}}{r_{12}},
\end{align}
where $w_{12}$ and $w'_{12}$ are the Cartesian components of $\mathbf{r}_{12}$.
In addition, a number of one-electron integrals are available (e.g., the dipole and higher-order multipole, hyperfine coupling, effective-core potential (ECP) integrals including
the spin--orbit part).
The standard molecular integrals are also implemented for gauge-including atomic orbitals.\cite{Reynolds2015PCCP}
The APIs for these integral programs are solely based on ``shell" objects (except for nuclear attraction potential, which requires a ``molecule" object as well),
which can be easily constructed externally.
Therefore, these integral routines can be interfaced to other program packages if desired.

\subsection{Code generator SMITH3}
Some of the components in BAGEL, including the CASPT2 nuclear gradient code, are implemented using the code generator SMITH3.\cite{smith}
SMITH3 can be thought of as a third-generation successor of the TCE code developed by Hirata in the 2000s,\cite{Hirata2003JPCA} replacing SMITH1 (Ref.~\onlinecite{Shiozaki2008PCCP})
and SMITH2 (unpublished) written by the author.
SMITH3 translates second-qauntized equations to tensor expressions, and then to compilable parallel C++ code.
It can handle both spin-free and spin-orbital expressions with multi-configurational reference functions, $|\Psi\rangle$, that are
written in the following forms: 
\begin{align}
&D_\Omega = \langle \Psi| \hat{E}^\dagger_\Omega \hat{{G}} \hat{{T}}|\Psi \rangle, \quad
D_\Omega = \langle \Psi | \hat{{T}}^{\prime\dagger} \hat{E}_\Omega \hat{{T}} | \Psi\rangle, \nonumber\\
&D_I = \langle I | \hat{{T}}^{\prime\dagger} \hat{{G}} \hat{{T}} | \Psi\rangle,
\end{align}
where $I$ is a Slater determinant, $\hat{E}_\Omega$ is the standard excitation operator,
and $\hat{G}$ and $\hat{T}$ are any general and excitation operators, respectively.
The tensors on the left-hand side are the target tensors.
All of the input and target tensors are kept in memory using the tiling algorithm as in the original TCE program.\cite{Hirata2003JPCA}
Getting, putting, and accumulating these tiles across nodes are implemented by means of
the MPI3 remote memory access protocol that allows us to perform efficient one-sided communication. 

\section{Developer contributions}
{\bf Toru Shiozaki} (TS) has designed the density fitting utilities and implemented as a parallel program.
He also implemented most of the molecular integrals and the non-relativistic HF, MP2, CASSCF, CASPT2, NEVPT2, and MRCI programs.
He is also responsible for the code for HF, MP2, CASSCF, and SA-CASSCF nuclear gradients and the {\it Z}-vector equations for HF and CASSCF reference functions.
The determinantal FCI code has been developed by {\bf TS} and {\bf Shane M. Parker} (SMP).
The parallel FCI is due to {\bf TS}.
All of the code for restricted active spaces is due to {\bf SMP}.
All of the FMM capabilities and ECP integrals were implemented by {\bf Hai-Anh Le} (HAL). Douglas--Kroll--Hess integrals were implemented by {\bf Yiqun Wang} (YW).

The code for state-specific CASPT2 nuclear energy gradients was developed by {\bf Matthew K. MacLeod} (MKM) and {\bf TS}.
The contributions by {\bf MKM} included the extension of SMITH3 to enable computation of the source terms in the {\it Z}-CASSCF equations.
This code was later extended by {\bf Bess Vlaisavljevich} (BV) to XMS-CASPT2 nuclear energy gradients and by {\bf Jae Woo Park} (JWP) to include the computation of derivative couplings.
Note that {\bf JWP} has significantly improved the CASPT2 gradient algorithm, which is currently used in the released version.
{\bf BV} implemented the numerical differentiation programs for nuclear Hessian elements and vibrational analyses.

The Dirac--Hartree--Fock program was developed by {\bf TS} with contributions from {\bf Matthew S. Kelley}.
The relativistic FCI was written by {\bf TS}. The initial relativistic CASSCF program was implemented by {\bf Jefferson E. Bates}.
The program has since been replaced by a more stable second-order algorithm by {\bf TS}, who also implemented the relativistic CASPT2 and MRCI programs.
{\bf Ryan D. Reynolds} (RDR) extended the relativistic CASPT2 and MRCI code to enable multi-state variants.
{\bf RDR} also developed a module that extracts EPR Hamiltonian parameters from the relativistic computations.
The pilot implementation of nuclear gradients within the relativistic framework was written by {\bf TS}.

{\bf JWP} implemented all of the geometry optimizers (for equilibrium geometries, transition states, and conical intersections)
that are currently used, replacing inefficient code due to {\bf TS}. {\bf JWP} is also responsible for the interfaces to dynamics programs.
All of the code for the ASD algorithms was developed by {\bf SMP} (except orbital optimization part written by {\bf Inkoo Kim}). The GIAO code was written by {\bf RDR}. 

{\bf Peter J. Cherry}, {\bf JWP}, {\bf BV}, {\bf HAL}, {\bf RDR}, {\bf Yeonjun Jeong}, {\bf Jheng-Wei Li}, {\bf YW} and {\bf TS}
wrote the manual. 

\section{Conclusion}
BAGEL is a highly efficient, full-fledged electronic structure package that is designed specifically for modern parallel computer hardware.
It implements unique features, such as CASPT2 nuclear energy gradients and relativistic multireference wave function methods, in addition to standard algorithms.
Every component in BAGEL is parallelized using both threads and MPI processes.
The program is distributed openly under the GNU General Public License Version 3+ at http://www.nubakery.org,
which allows users and theory developers to freely download, modify, and redistribute it under the same license.

\section{Further reading}
The details of the input syntax and available options, as well as references to the underlying methods can be found in the BAGEL Manual. The BAGEL Manual is available at
http://www.nubakery.org.

\begin{acknowledgments}
All of the developers described in the above section have greatly contributed to the BAGEL program package, to whom I am grateful.
I am also indebted to Takeshi Yanai for his help on the second-order CASSCF implementation and for numerous stimulating conversations.
Jeff Hammond is thanked for introducing to us the MPI3 remote memory access protocol.
Since 2012, the development of BAGEL has been supported by US Department of Energy, Basic Energy Sciences (DE-FG02-13ER16398),
Air Force Office of Scientific Research Young Investigator Program (FA9550-15-1-0031),
the National Science Foundation CAREER Award (CHE-1351598), and National Science Foundation SI2-SSI (ACI-1550481).
The author is an Alfred P. Sloan Research Fellow.
\end{acknowledgments}

\end{document}